\documentclass[aps, 10pt, twocolumn, showpacs, superscriptaddress]{revtex4-2}

\usepackage[utf8]{inputenc}
\usepackage{amsmath,amssymb,amsfonts,amsthm,amsopn,amscd,bm}
\usepackage{dsfont}
\usepackage{multirow}
\usepackage{graphicx}
\usepackage{booktabs}
\usepackage[shortlabels]{enumitem}
\usepackage{comment}
\usepackage{xcolor}
\usepackage{enumitem}
\usepackage[normalem]{ulem}

\let \vec \textbf

\usepackage[breaklinks]{hyperref}
\hypersetup{colorlinks,
   linkcolor={red!50!black},
    citecolor={blue!50!black},
    urlcolor={blue!80!black}
}

\usepackage{orcidlink}

\DeclareMathOperator{\tr}{tr}
\DeclareMathOperator{\argmax}{argmax}

\newcommand{\bra}[1]{\mathinner{\langle #1|}}
\newcommand{\ket}[1]{\mathinner{|#1\rangle}}

\newcommand{\ketbra}[2]{\mathinner{|#1\rangle\langle#2|}}

\newcommand{\nn}{\nonumber}

\renewcommand{\AA}{\mathcal{A}}
\newcommand{\BB}{\mathcal{B}}
\newcommand{\CC}{\mathcal{C}}
\newcommand{\DD}{\mathcal{D}}

\newcommand{\II}{\mathcal{I}}
\newcommand{\MM}{\mathcal{M}}

\newlength{\minuslength}
\begin{document}

\title{Bell inequalities with overlapping measurements}

\author{Moisés Bermejo Morán${}^{\orcidlink{0000-0003-1441-0468}}$}
\email{moises.moran@uj.edu.pl}
\affiliation{
Faculty of Physics, Astronomy and Applied Computer Science, Institute of Theoretical Physics, Jagiellonian University,
30-348 Krak\'{o}w, 
Poland}

\author{Alejandro Pozas-Kerstjens${}^{\orcidlink{0000-0002-3853-3545}}$}
\affiliation{
Institute of Mathematical Sciences (CSIC-UAM-UC3M-UCM), 28049 Madrid, Spain}

\author{Felix Huber${}^{\orcidlink{0000-0002-3856-4018}}$}
\affiliation{
Faculty of Physics, Astronomy and Applied Computer Science, Institute of Theoretical Physics, Jagiellonian University,
30-348 Krak\'{o}w, 
Poland}

\date{\today}

\begin{abstract}
Which nonlocal correlations can be obtained, when a party has access to more than one subsystem?
While traditionally nonlocality deals with spacelike separated parties, this question becomes important with quantum technologies that connect devices by means of small shared systems.
Here we study Bell inequalities where measurements of different parties can have overlap. This allows to accommodate problems in quantum information such as the existence of quantum error correction codes in the framework of non-locality.
The scenarios considered show an interesting
behaviour with respect to Hilbert space dimension, overlap, and symmetry.
\end{abstract}

\maketitle

The relation between the whole and its parts is a concept central to quantum many-body physics,
as it determines what correlations, and thus emergent physical phenomena, a system can exhibit.
To understand this relation, the {\em quantum marginal problem} has played a key role.
The simplest instance of
this QMA-complete
quantum constraint satisfaction problem
asks~\cite{Liu2006}:
given the reduced density matrices $\varrho_{AB}$, $\varrho_{AC}$, and $\varrho_{BC}$, does there exist a joint state $\varrho_{ABC}$ of which these are its marginals?
Finding the answer to this question is not only relevant to determine the ground state energy of local Hamiltonians~\cite{RevModPhys.35.668,https://doi.org/10.48550/arxiv.2212.03014}, but also in the study of multipartite entanglement:
for example, an interesting question is the existence 
of states which achieve maximal entanglement in every bipartition \cite{HIGUCHI2000213, huber2018ame, PhysRevLett.128.080507}.

In this type of problems, 
the dependence on
the local system sizes
and on the overlap of the collections of subsystems involved makes them equally challenging and intriguing.
The main goal of this work is to formulate analogous questions in Bell nonlocality: 
what correlations can be obtained in a multipartite quantum system, 
by players that are allowed to measure overlapping collections of subsystems? 
Does this scenario present advantages over measuring only locally, and how does the maximum quantum value change with the local systems and overlap sizes?
In the current status of quantum technologies, where scaling-up plans consider connecting devices by means of small shared systems \cite{bravyi2016trading,eddins2022double,piveteau2022knitting}, the questions posed above acquire particular relevance in the context of assessing the quality of devices in a device-independent manner.

\begin{figure}[tbp]
    \centering
    \includegraphics[width=0.95\columnwidth]{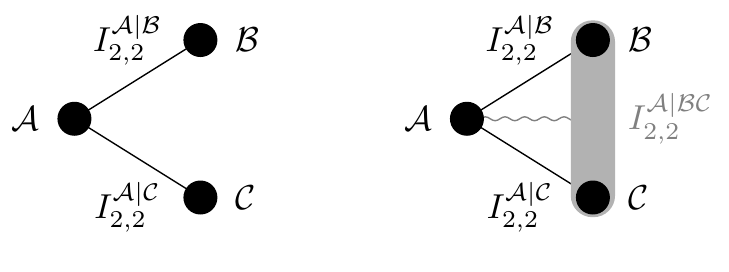}
    \caption{
    (Left) Scenario 1. The maximal quantum value of the joint Bell inequality $J_2 = I_{2,2}^{\AA|\BB} 
    + I_{2,2}^{\AA|\CC}$ in Eq.~\eqref{eq:BI}
    is governed by size of local physical systems. 
    (Right) Scenario 2. In the overlapping Bell inequality 
    $K_2 =  I_{2,2}^{\AA|\BB} 
    + I_{2,2}^{\AA|\CC}
    + I_{2,2}^{\AA|\BB\CC}$
    in Eq.~\eqref{eq:OBI}
    a fourth party has,
    independently of the measurements performed on Bob and Charlie, also access to measurements on the joint system.
    Here both the dimensions of the individual subsystems as well as the overlap of $\BB\CC$ with $\BB$ and $\CC$ affect the maximal quantum value.
    \label{fig:JK}}
\end{figure}

For instance, take three parties, $\AA$, $\BB$ and $\CC$, that attempt to simultaneously maximize the value of two CHSH inequalities between $\AA$ and $\BB$, and between $\AA$ and $\CC$ 
(Fig.~\ref{fig:JK}, left).
If party $\AA$ has a system of local dimension~2, 
monogamy of entanglement prohibits a simultaneous maximal quantum value \cite{toner2006monogamy}. 
Party $\AA$ having a ququart allows for sharing two maximally entangled states (with $\BB$ and $\CC$), while a qutrit will interpolate these scenarios. Additional complexities arise if a third party has,  independently of the measurements performed on $\BB$ and $\CC$, also access to measurements on the joint $\BB\CC$ system
(Fig.~\ref{fig:JK}, right). 
In line with recent efforts~\cite{frerot2023review,muller2021mbbi}, we here treat maximal quantum values of Bell inequalities purely as a measure of quantum correlations:
understanding how the highest quantum values can be extracted in overlapping scenarios, as if done by a single experimentalist in a single laboratory, is our aim.

In this work we put forward questions on quantum nonlocality from the perspective of the quantum marginal problem and we provide numerical methods to address some of them.
We exemplify this in two concrete scenarios, but the methods develop are applicable more broadly.
We find several interesting features, 
such as an separable-entangled-separable transition in optimal strategies when the local dimension is changed, 
as well as a ``symmetry-breaking'' phenomenon for Bell inequalities with overlapping subsystems.

\section{Setting}
\label{sec:setting}

Consider a scenario where Alice, Bob and Charlie share a tripartite quantum system. 
A fourth person, Dave, has access 
to the joint system of Bob and Charlie.
What quantum correlations can they obtain by making measurements?
%
Consider the following linear combinations of Bell inequalities,
\begin{align}
J_d  &=  I_{2,d}^{\AA|\BB} 
    + I_{2,d}^{\AA|\CC}\,, \label{eq:BI} \\
   K_d &=  I_{2,d}^{\AA|\BB} 
    + I_{2,d}^{\AA|\CC}
    + I_{2,d}^{\AA|\BB\CC}\,, \label{eq:OBI}
\end{align}
where $I_{2,d}^{S|T}$ is the SATWAP inequality 
involving two measurements with $d$ outcomes
between subsystems $S$ and~$T$~\cite{salavrakos2017Bell}.
The SATWAP inequality self-tests for maximal $d$-level entanglement, where it can achieve its maximum value of $2(d-1)$
(more details in Appendix~\ref{app:satwap}).

In Eq.~\eqref{eq:BI}, 
Alice has access to four measurements in total:  
two measurements relevant to the inequality with Bob, $I_{2,d}^{\AA|\BB}$, and 
two for the inequality with Charlie, $I_{2,d}^{\AA|\CC}$, 
all acting on the same physical system. 
Bob and Charlie have also access to two measurements.
In the more complicated scenario $K_d$ corresponding to Eq.~\eqref{eq:OBI}, 
Alice has access to three pairs of measurements: one for each of the partitions $\AA|\BB$, $\AA|\CC$ and $\AA|\BB\CC$ 
\footnote{
One could also consider a minus sign in the last term of Eq.~\eqref{eq:OBI}. 
Numerically we observe the same behaviour, thus we restrict our discussion to Eq.~\eqref{eq:OBI}
}. 
Bob and Charlie each has access to a pair of measurements on their subsystem. Here we additionally consider also a fourth party, Dave, that has access to another pair of measurements on the {\it joint} system of Bob and Charlie. 

The key difference between Eq.~\eqref{eq:BI} and Eq.~\eqref{eq:OBI} is that Dave's measurements overlap with those of Bob and Charlie. 
In contrast to regular Bell inequalities as appearing in Eq.~\eqref{eq:BI}, 
we call expressions as in Eq.~\eqref{eq:OBI} 
{\it overlapping Bell inequalities}. 
Graphically, we depict these overlapping Bell inequalities with wiggly lines as done in 
Figs.~\ref{fig:JK} (right) and \ref{fig:merge}.
In these scenarios, the assistance of Dave could allow Alice, Bob and Charlie to achieve additional nonlocal correlations; a feature that could serve to test current quantum devices with access to shared systems.
 
In the case that the local Hilbert space dimension is unconstrained, both \eqref{eq:BI} and \eqref{eq:OBI} can achieve their maximum quantum value
by taking tensor products of $d$-level maximally entangled states.
However, if the local dimension is sufficiently small, a type of frustration appears: 
not all terms can be simultaneously maximized.
A particularly interesting example is to consider two outcomes, 
where one can expect that if Alice's system has dimension two, it exhibits maximal frustration,
dimension four allows for a tensor-product strategy,
and dimension three interpolates these scenarios.

\section{Methods}
\label{sec:methods}

What methods allow to find the maximum of Eqs.~\eqref{eq:BI} and \eqref{eq:OBI} for finite dimensional quantum states?
There are two main obstacles.
The first obstacle originates from the dimensional constraints, where the usual noncommutative polynomial optimization methods \cite{pironio2010convergent} are not directly applicable.
The second obstacle arises from the overlap between some subsystems, 
namely those of Bob, Charlie, and Dave.
%

\emph{Scenario 1: no overlap}. the inequality $J_d$ in Eq.~\eqref{eq:BI} has no overlap, and all the constraints on the distribution of nonlocal correlations come from the local \emph{dimension} of Alice's system. 
The maximum value over quantum correlations can be formulated as an optimization problem,
\begin{align}\label{eq:BIopt}
 \max \tr\big(\varrho \, J_d(A, B, C)\big)
\end{align}
subject to the commutativity constraints
\begin{align}\label{eq:BI_commconst}
[A_{a|x}^{\AA|\BB}, B_{b|y}^{\AA|\BB}] & = 0, & [A_{a|x}^{\AA|\CC}, B_{b|y}^{\AA|\BB}] & = 0, \nn\\
[A_{a|x}^{\AA|\BB}, C_{c|z}^{\AA|\CC}] & = 0, & [A_{a|x}^{\AA|\CC}, C_{c|z}^{\AA|\CC}] & = 0, \nn\\
[B_{b|y}^{\AA|\BB}, C_{c|z}^{\AA|\CC}] & = 0, &
\end{align}
where $J_d(A, B, C) = J_d(A^{\AA|\BB}, A^{\AA|\CC}, B^{\AA|\BB}, C^{\AA|\CC})$.
The maximization in Eq.~\eqref{eq:BIopt} runs over all states $\varrho$ in some Hilbert space $\mathcal H_\AA \otimes \mathcal H_\BB \otimes \mathcal H_\CC$ of the given local dimensions and the $d$-outcome measurements
\begin{align}
& A^{\AA|\BB}_1, \quad A^{\AA|\BB}_2, 
\quad A^{\AA|\CC}_1, \quad A^{\AA|\CC}_2,\nn\\
& B^{\AA|\BB}_1, \quad B^{\AA|\BB}_2, 
\quad C^{\AA|\CC}_1, \quad C^{\AA|\CC}_2. 
\end{align}
We denote the elements of $A^{\AA|\BB}_1$ (the effects) as
\begin{equation}
    A^{\AA|\BB}_1 = \left\{ A^{\AA|\BB}_{1|1}, \ldots,  A^{\AA|\BB}_{d|1}\right\},
\end{equation}
following the convention, and similarly for the other measurements.
The elements of each measurement must be positive semidefinite and add up to the identity.  
The notation is chosen so that the label of each measurement indicates the party performing it, the superscript denotes the partition in which the measurement is involved, and the subscript enumerates all such measurements.

Every measurement with two effects can be obtained as the convex combination of projective measurements \cite[Lemma 3]{PhysRevLett.119.190501}. 
Thus, for inequalities involving measurements with two effects as Eqs.~\eqref{eq:BI} and \eqref{eq:OBI}, we can assume the effects to be projectors.
We will exploit this equivalence in the numerical implementation, since each technique is more suited to a different assumption on the effects.

The optimal solution of Eq.~\eqref{eq:BIopt} is generally not straightforward to compute, but we can obtain numerical bounds. 
For fixed local dimensions, the see-saw algorithm \cite{PhysRevA.64.032112} provides lower bounds 
by numerically optimizing over 
states and observables.
The algorithm starts with random initial state and effects,
and 
one then alternates between optimizing over 
the state and over the effects. 
To obtain upper bounds, we use the {\em sampling-based} moment relaxations proposed in~\cite{navascues2015characterizing}.
Moment relaxations are frequently used to obtain outer approximations to polynomial optimization problems \cite{pironio2010convergent}. 
To impose dimensional constraints, one can {\em sample} a basis of feasible moments in the fixed dimension. A detailed description is provided in Appendix \ref{app:bounds}.

{\emph{Scenario 2: overlap}. Similarly, the maximal quantum value for the overlapping Bell inequality $K_d$ in Eq.~\eqref{eq:OBI} between Alice, Bob, Charlie and Dave, is the solution of the optimization problem
\begin{align}
\label{eq:OBIopt}
\max \tr\big(\varrho K_d(A, B, C, D) \big)
\end{align}
subject to the commutativity constraints
\begin{align}\label{eq:OBI_commconst}
[A_{a|x}^{\AA|\BB\CC}, B_{b|y}^{\AA|\BB}] & = 0, &  [A_{a|x}^{\AA|\BB}, D_{d|w}^{\AA|\BB\CC}] & = 0, \nn\\
[A_{a|x}^{\AA|\BB\CC}, C_{c|z}^{\AA|\CC}] &= 0,  & [A_{a|x}^{\AA|\CC}, D_{d|w}^{\AA|\BB\CC}] &= 0, \nn\\
[A_{a|x}^{\AA|\BB\CC}, D_{d|w}^{\AA|\BB\CC}] &= 0, &
\end{align}
in addition to those of Eq.~\eqref{eq:BI_commconst}.

The optimization in~\eqref{eq:OBIopt} runs over two additional pairs of $d$-outcome measurements
\begin{align}
A^{\AA|\BB\CC}_1, \quad A^{\AA|\BB\CC}_2, \quad
D^{\AA|\BB\CC}_1, \quad D^{\AA|\BB\CC}_2.
\end{align}
The constraints in~\eqref{eq:OBI_commconst} express that measurements performed over nonoverlapping subsystems commute.
The main difference here is that Dave's measurements, 
$D_j^{\AA|\BB\CC}$,
are not required to commute with those of Bob or Charlie. 
Note that it is not possible to assign a joint probability distribution for this scenario; and traditional tools to give upper bounds on Bell inequalities are limited.

The tools we present below produce upper bounds by relaxing the setting. There are several ways these scenarios involving Bell inequalities can be relaxed.
Namely, (a) splitting terms in the joint inequality, (b) increasing the local dimensions, and (c) dropping commutativity constraints. The seesaw algorithm still provides dimension-specific lower bounds for Eq.~\eqref{eq:OBIopt} that match the optimal value in the scenarios for which this can be computed.

\begin{figure*}[tbp]
\centering
\includegraphics[width=0.8\textwidth]{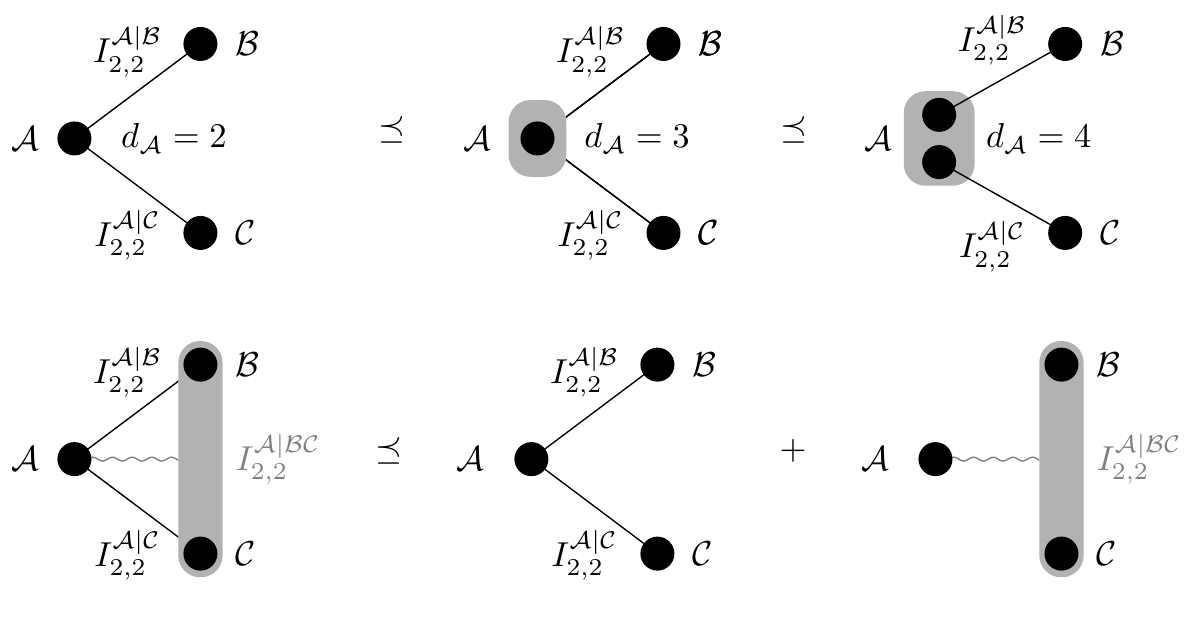}
\caption{{\bf Bell inequalities with overlap.} 
Above: Bell inequality $J_2$ in Eq.~\eqref{eq:BI}, where only the local dimensions play a role. Bellow: overlapping Bell inequality $K_2$ in Eq.~\eqref{eq:OBI}, where both the local dimensions and the overlapping structure play a role.}
\label{fig:merge}
\end{figure*}

\section{Results}
\label{sec:results}


We consider Eqs.~\eqref{eq:BI} and~\eqref{eq:OBI} for $d=2$, corresponding to two outcomes. Here one expects that a two-dimensional system of Alice will show maximal frustration, a four-dimensional system allows for a tensor-product strategy, and a three-dimensional system interpolates these scenarios.

Table \ref{tab:chains} shows upper and lower bounds obtained with the methods from Section~\ref{sec:methods}. The lower bounds are obtained with the see-saw algorithm for general positive effects.
We provide the optimal states and measurements in Appendix \ref{app:optJ2}.
In contrast, the upper bounds are obtained with the moment relaxations and assume the effects to be projectors, since this drastically reduces the size of the problem. 
We see bounds arising from two different sources: the finite dimension and  the overlap between measurements. 

\smallskip
\paragraph*{Scenario 1: no overlap.}
Let us discuss the inequality $J_2$ first, as here already effects from the finite dimension play a role.
The setting $J_{2}$ in Eq.~\eqref{eq:BI} consists of two joint $I_{2,2}$ inequalities between $\AA|\BB$ and $\AA|\CC$, each involving two dichotomic measurements per party~\footnote{Note the difference to the setting of \cite{toner2006monogamy}, where Alice uses the same measurements with both Bob and Charlie.}. 
In this setting, increasing the dimension of the system gives a higher quantum value, thus we can use these correlations to test the dimension of $\AA$.

\begin{table}[tbp]
   \centering
   \begin{tabular}{@{}cccccc@{}} 
     Inequality & $\operatorname{c + q}$ & $\operatorname{q + q}$ & $(d_A, d_B, d_C)$ & $\operatorname{lb}$ & $\operatorname{ub}$ \\
     \toprule
      & & & $(2,2,2)$ & 3.4142 & \ $3.4142$ \\
     $J_2$ & $2+\sqrt{2}$ & 4 & $(3,2,2)$ & 3.6365 & \ $3.6365$ \\
      & & & $(4,2,2)$ & 4.0000 & \ $4.0000$ \\
     \midrule
      & & & $(2,2,2)$ & 5.4142 & \ $5.4142$ \\
     $K_2$ & $4+\sqrt{2}$ & 6 & $(3,2,2)$ & 5.5096 &  \ $5.6365$ \\
      & & & $(4,2,2)$ & 6.0000 & \ $6.0000$ \\
      \bottomrule
   \end{tabular}
   \caption{{\bf Bounds on $J_{2}$ and $K_2$.}
   Systems with different local dimensions $d_A$, $d_B$ and $d_C$ obtain different maximal quantum values in $J_{2}$ [Eq.~\eqref{eq:BI}] and $K_2$ [Eq.~\eqref{eq:OBI}]. 
   The value $\operatorname{q+q}$ corresponds to the case where bipartite Bell inequalities individually achieve their quantum maximum, and $\operatorname{c+q}$ to the case where one inequality achieves classical maximum only. Lower bounds (lb) are obtained with see-saw algorithms. Upper bounds (ub) are computed via moment matrix sampling for the case of $J_2$. In the case of $K_2$ the upper bounds are computed by adding the corresponding value of $J_2$ and the maximum value of the additional SATWAP inequality. 
   Changing the sign in the last term of $K_2$
   yields the same values.
   \label{tab:chains}}
\end{table}

When Alice has a two dimensional system, the maximum value is $2 + \sqrt 2 \approx 3.4142$. This value corresponds to the maximal quantum value in $\AA|\BB$ and the maximal classical value in $\AA|\CC$.
Moreover, this is the quantum maximum since it saturates the upper bound obtained via moment relaxations. 
%
When Alice has a four-dimensional system, she can share a maximally entangled qubit with both Bob and Charlie, and thus achieve the quantum maximum for both inequalities simultaneously, $4 = 2 + 2$.

A three-level system on Alice's side interpolates these two cases:
the maximum must be strictly smaller than $4$, since there is not enough physical space available to hold two qubits. However, it is not clear whether a higher value can be obtained with qutrits than with qubits. Numerically we see that this is the case, so the correlations in this setting can be used to test the physical dimension that Alice has access to.
The lower bounds from the see-saw algorithm and the upper bounds from the moment relaxations meet at a value of $3.636(5)= 2\times 1.818(3)$. Note that the optimal strategy in dimension three involves a tripartite entangled state, while those for dimensions two and four only need bipartite maximally entangled states.

\smallskip
\paragraph*{Scenario 2: overlap.}
Now consider the  Bell inequality $K_{2}$ from Eq.~\eqref{eq:OBI}
where a fourth party has access to joint measurements on $\BB\CC$.
Comparing the maximum values achievable for this operator and for $J_2$ one can discern whether having access to joint measurements on the $\BB\CC$ system allows to demonstrate more nonlocality.

When Alice holds a two- or a four-dimensional system, the maximal values achievable can be directly derived from the results for $J_2$ (note that $K_2 = J_2 + I^{\AA|\BB\CC}_{2,2}$), just by reusing the measurements in $\AA|\BB$ for $\AA|\BB\CC$. Since these measurements achieve the maximum quantum value of the corresponding SATWAP inequality, the value of $K_2$ achieved is the maximum quantum value.

Again, the scenario is more interesting when the local dimension of Alice is three. Following the prescription above one would obtain a value of $1.8183$ for each of the SATWAP inequalities, and thus a maximal value for $K_2$ of $3 \times 1.818(3) = 5.454(8)$.
However, using the see-saw one finds the higher value of $5.509(6) = 1.683(7) + 1.913(0) + 1.913(0)$.

Interestingly, for dimension three on Alice's side the strategy maximizing $K_2$ seems to favour asymmetric correlations, such that e.g. $\AA|\BB$ achieves a higher value than $\AA|\CC$, and reuses the measurements  for $\AA|\BB\CC$. 
This asymmetric strategy is however not optimal for $J_2$, since these measurements only achieve a value of $1.683(7)+1.913(0) = 3.596(7) < 3.636(5)$. 
Indeed, numerical results suggest optimal strategies for $K_2$ come from optimal strategies of either $2 I_{2,2}^{\AA|\BB} + I_{2,2}^{\AA|\CC}$ or $ I_{2,2}^{\AA|\BB} + 2I_{2,2}^{\AA|\CC}$.

Can we obtain the maximum in both  $J_2$ and $K_2$ by using the same strategy? It seems that this is not the case:
using see-saw, the maximum value of $K_2$ can not be achieved, 
when fixing the value of the term $J_2$ to its maximum. 
Optimising $K_2$ constrained to keep the maximal value for $J_2$, we get
\begin{equation}
5.456(6) = 1.816(5) + 1.820(0) + 1.820(0)
\end{equation}
and then $K_2$ seems to gain no advantage by measuring the joint system,
as the value of $I^{\AA|\BB\CC}$ can be obtained by reusing the measurements used in $I^{\AA|\CC}$. 

Lastly, we note that there are radical changes in the
optima when changing the local systems dimensions. The
optimum in $J_2$ of $2+\sqrt 2$ can not be achieved while fixing
the first term, $I^{\AA|\BB}_{2,2}$, to be $(2 + \sqrt
2)/2$. Numerical experiments with moment relaxations suggest that an arbitrarily close to optimal quantum value between Alice
and Bob limits Alice and Charlie to attain classical Bell
values, and vice versa.
Our conclusion thus is that, there are symmetrical optimal strategies for $J_2$, while optimal strategies for $K_2$
seem to be asymmetric.

\section{Conclusions}
Multipartite quantum systems display a rich behaviour in terms of their correlations~\cite{https://doi.org/10.48550/arxiv.2302.03139, walter2016multipartite, PhysRevA.107.022412}.
This work shows that such rich behaviour is also present at the level of nonlocality,
where the landscape of quantum correlations becomes even more intriguing when parties have joint access to multiple subsystems.
Additionally, as with nonoverlapping scenarios, the local systems size plays a role in the ability to distribute nonlocal correlations amongst many parties.

A variation of our Scenario 1 depicted in (Fig.~\ref{fig:JK}, left) has been studied in the context of monogamy of correlations \cite{toner2006monogamy}. 
There, Alice was restricted to use the same measurements for both Bob and Charlie. 
Under this additional restriction it is not possible to separate classical from nonclassical correlations with simple inequalities such as $J_2$ in Eq.~\eqref{eq:BI}. Our work shows that, if this restriction is lifted, a gap appears, and with it potential quantum advantages and means of witnessing the dimension of the underlying quantum states.

The results above deal with a small number of systems for illustration purposes, but the methods developed can be, in principle, used to tackle important questions in quantum information.
An interesting connection is that to quantum codes: the settings considered here can be seen as the equivalent to entanglement distributions seen in quantum codes, but expressed in terms of nonlocality. 
For a pure code of distance $d$ and block length $n$, 
one requires that maximal entanglement can be recovered between every subsystem of size $(n-d+1)$ with a reference system, illustrated in Fig.~\ref{fig:qecc}~\cite{Huber2020quantumcodesof}.
In particular, the existence of so-called absolutely maximally entangled states can be tested by finding upper bounds on the quantum value of overlapping Bell inequalities on multipartite systems of finite dimensions. 
Another similarity of our setting is to random access codes, where one wants to encode information into a subsystem that is too small to contain it~\cite{PhysRevLett.113.100401}.

\begin{figure}[tbp]
    \includegraphics{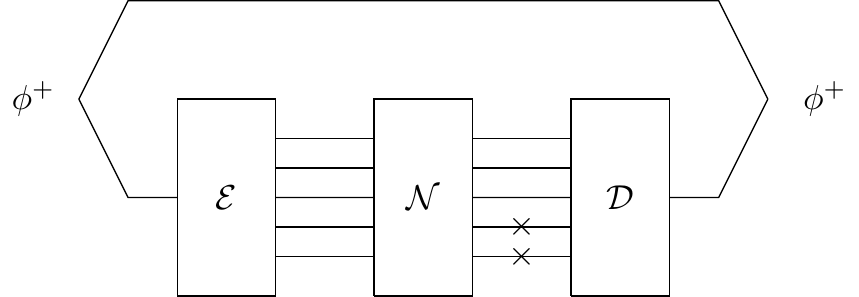}
\caption{\textbf{Quantum error correction code}. 
A code has distance $d$, if maximal entanglement can be recovered between a reference system and every code subsystem of size $n-d+~1$~. This can be characterized in terms of the perfect recovery of a maximally entangled state $\phi^+$, one part of which is encoded, acted upon by noise, and decoded.}
\label{fig:qecc}
\end{figure}

In order to be able to properly attack these problems, it is important to develop tractable ways of characterizing overlapping Bell scenarios,
for example via moment relaxations in the spirit of the Navascu\'es-Pironio-Ac\'in hierarchy \cite{navascues2008convergent}.
Here it is unclear how a joint probability distribution for these type of scenarios can be defined.
Also, it would be interesting to understand how optimal strategies involving overlapping violate the symmetry of the set-up, and how the separable-entangled-separable transition of Scenario 1 (Fig. \ref{fig:JK}, left) with changing dimension of Alice generalizes for higher dimensional systems. 
Lastly, it would be interesting to understand whether connections to frustrated ground states of quantum many-body Hamiltonians can be made~\cite{PhysRevA.88.020301}.

{\it Acknowledgements.}
We thank Armin Tavakoli for feedback on the manuscript and Paweł Horodecki for fruitful discussions.
MBM and FH were supported by the FNP through TEAM-NET (POIR.04.04.00-00-17C1/18-00).
APK's work is supported by the Spanish Ministry of Science and Innovation MCIN/AEI/10.13039/501100011033 (CEX2019-000904-S and PID2020-113523GB-I00), the Spanish Ministry of Economic Affairs and Digital Transformation (project QUANTUM ENIA, as part of the Recovery, Transformation and Resilience Plan, funded by EU program NextGenerationEU), Comunidad de Madrid (QUITEMAD-CM P2018/TCS-4342), the CSIC Quantum Technologies Platform PTI-001 and Universidad Complutense de Madrid (FEI-EU-22-06).

\appendix

\section{SATWAP inequality}
\label{app:satwap}

A generalization of the the celebrated CHSH inequality
$\langle A_0 B_0 \rangle + \langle A_1 B_0 \rangle - \langle A_0 B_1 \rangle + \langle A_1 B_1\rangle$,
is the Salavrakos-Augusiak-Tura-Wittek-Acin-Pironio  (SATWAP) inequality~\cite{salavrakos2017Bell}. It is a bipartite inequality involving $m$ measurements per party, each with $d$ possible outcomes. The expression is given by
\begin{equation}
    I_{m,d} = \sum_{k = 0}^{\lfloor d/2 \rfloor - 1} 
    ( \alpha_k \mathds P_k - \beta_k \mathds Q_k ),
\end{equation}
where 
\begin{align}
\mathds P_k 
&= \sum_{i = 1}^m \Big(p(A_i = B_i + k) + p(B_i = A_{i+1} + k)\Big) \nonumber\,,\allowdisplaybreaks\\
\mathds Q_k 
&= \sum_{i = 1}^m \Big(p(A_i = B_i - k -1) + p(B_i = A_{i+1} - k - 1)\Big)\nonumber\,,
\end{align}

and the parameters $\alpha_k$ and $\beta_k$ are given by
\begin{equation}
    \begin{split}
        \alpha_k & = \frac{1}{2d}\tan\left( \frac{\pi}{2m}\right)\left[g(k) - g\left( \left\lfloor \frac{d}{2} \right\rfloor\right) \right],\\
        \beta_k & =  \frac{1}{2d}\tan\left( \frac{\pi}{2m}\right)\left[g(k + 1 - \frac{1}{m}) + g\left( \left\lfloor \frac{d}{2} \right\rfloor\right) \right],
    \end{split}
    \end{equation}
where $g(x) = \cot(\pi(x + 1/2m)/d)$. Its maximum over latent hidden variable (i.e., classical) models is 
\begin{equation}
    (1/2)\tan(\pi/2m)[(2m - 1)g(0) - g(1 - 1/m)] - m\,,
\end{equation}
while over quantum models it is $m(d-1)$.
Interestingly, the SATWAP inequality is maximized by systems with local dimension $d$, and in such a case it self-tests for the presence of maximal entanglement.

\section{Numerical bounds}
\label{app:bounds}
The maximal quantum value in settings such as those in Eqs.~\eqref{eq:BI} and \eqref{eq:OBI} are generally not straightforward to compute.
Thus we aim to find bounds from problems that are easier to solve. The first method we discuss is nonnumerical, it is based on a direct comparison between optimization problems. The second and third one are numerical algorithms to provide respectively lower and upper bounds to the optimization problems in Eqs.~\eqref{eq:BIopt} and \eqref{eq:OBIopt}.

\smallskip
\noindent{\bf Comparing problems.}
\label{subsec:comparing}
Note that by relaxing an optimization problem, its feasible region becomes larger.
Consequently, also a higher objective value can be obtained, leading to upper bounds.
There are several ways Bell inequalities such as Eqs.~\eqref{eq:BI} and \eqref{eq:OBI} can be relaxed: 
first, the maximum of a joint inequality can not be higher than the sum of the separate maxima of its terms.
Second, increasing the system dimension gives access to a wider class of correlations.
Third, commuting measurements can be relaxed to partially overlapping measurements.
The converse reasoning leads to lower bounds 
from a strengthening of the constraints,
in addition to the 
lower bound corresponding to the classical maximum of the inequality, where all measurements commute.

\smallskip
\noindent{\bf Numerical lower bounds.}
\label{subsec:sewsaw}
For fixed local dimensions, the see-saw algorithm can provide lower bounds on Eqs.~\eqref{eq:BIopt} and \eqref{eq:OBIopt} by numerically optimizing over 
states and observables~\cite{PhysRevA.64.032112}.
The algorithm starts with random initial state and  measurements. 
One then alternates between optimizing over 
the state and over the measurements. 

More specifically, denote by $\varrho$ the state.
Then group the measurements into a set $X$ 
and its complement $Y$ 
(more generally more subsets can be used), 
in a way such that each term of the inequality 
involves one observable from $X$ and one observable from $Y$. 
For example, in Eq.~\eqref{eq:BIopt} we can take $X = A^{\AA|\BB}\cup A^{\AA|\CC}$ and $Y = B^{\AA|\BB} \cup C^{\AA|\CC}$.
Then do the following:
\begin{enumerate}
    \item Choose random $\varrho_0$, $X_0$, $Y_0$.
    \item $\varrho_{i+1} \,= 
     \argmax_{\varrho_i\,} \tr\big(\varrho_i \, J_d(X_i, Y_i)\big)$
    \item $X_{i+1} = \argmax_{X_i} \tr\big(\varrho_i \, J_d(X_i, Y_i)\big)$
    \item $Y_{i+1} \,\,= \argmax_{Y_i} \tr\big(\varrho_i \, J_d(X_i, Y_i)\big)$
    \item Repeat steps 2, 3 and 4 until convergence.
\end{enumerate}
The optimizations in the steps 2, 3, and 4 can be performed by a semidefinite program, 
which can be run on a desktop computer
for small matrix sizes. 
Note that there is no guarantee for the algorithm to converge to the global maximum. 
However, by running it many times one can often obtain reasonable lower bounds.

\smallskip
\noindent{\bf Numerical upper bounds.}
Fixing the local dimension presents a particular challenge for finding the maximal quantum value of a Bell inequality.
The methods by Ref.~\cite{navascues2015characterizing}
allow to obtain upper bounds 
through {\em sampling-based} moment relaxations. 
Moment relaxations are frequently used to obtain outer approximations to polynomial optimization problems \cite{pironio2010convergent}. 
For our purposes, the moments involved consist of expectation values of products of measurements, 
for example 
$\tr(\varrho A^{\AA|\BB}_{1|1} B^{\AA|\BB}_{1|1})$. 
To impose dimensional constraints, one can then {\em sample} a basis of feasible moments for a given fixed dimension. 
As we show below this strategy can also work for overlapping inequalities such as Eq.~\eqref{eq:OBIopt}. 

More specifically, this method works in the following way:
$N$ noncommutative Hermitian variables
$\vec x = (x_1, \ldots, x_N)$
generate a sequence $\mathcal I$ of monomials. For example, the sequence $\II$ of monomials up to degree two reads
\begin{equation}
\mathcal I = (1,x_1, \ldots, x_N, x_1^2, x_1 x_2, x_2 x_1, \dots,  x_N^2)\,.
\end{equation}
The associated moment matrix, indexed by the monomials $p$ and $q$ in $\mathcal I$, has entries
$
\MM_{pq} = \tr(\varrho\, p^\dag q)
$,
and is positive semidefinite for any $\varrho$.

To form an approximation to the set of 
finite-dimensional quantum correlations, 
we now consider the span of valid moment matrices $\MM$ \cite{navascues2015characterizing,navascues2015bounding}.
For this, one samples quantum states and measurements until one has obtained a basis of moment matrices. 
In practice, this means
that one samples moment matrices
until the span of the sampled moment matrices stabilizes.
This can be done for example by extending 
the current orthonormal set of moment matrices by a new matrix obtained through
Gram-Schmidt orthonormalization.

To detect when we complete the basis, 
we sample different rank classes of projectors separately.
This produces for each class a basis of feasible moment matrices corresponding to projective valued measurements. 
For each class, one then optimizes over the positive matrices the corresponding basis spans, which corresponds to solving a semidefinite program. 
As the optimum is obtained in some rank class, 
this suffices to obtain the maximum over all 
the classes and gives an upper bound for the 
optimum of the original problem.

In principle, for increasing indexing sequences of monomials this approach converges 
in the non-overlapping scenario to the optimum of a noncommutative polynomial 
in finite-dimensional matrix variables \cite{navascues2015characterizing}. 
However, the numerical precision required in the Gram-Schmidt orthogonalization process can often be too demanding
for a large number of variables and relaxation order.
Thus the main difficulties in this approach involve the sampling 
of a complete basis and solving the resulting relaxation. 
If the sequence $\II$ of monomials is too small, 
one may not be able to obtain good bounds from the relaxations. 
If the sequence is too large, 
numerical errors from the Gram-Schmidt orthonormalization
may dominate before one can complete a basis. 
Lastly, the resulting semidefinite program may simply be too large in size to solve on a standalone computer. 


These difficulties are directly related to the number of monomials in the indexing sequence $\II$.
Choosing a ``good" sequence $\II$ is thus an interesting problem. 
One way to approach this computational barrier is to exploit the symmetries of the setting.
This both reduces the number of rank classes
and can be used to symmetry-reduce the moment matrix~\cite{https://doi.org/10.48550/arxiv.2112.10803}. 
For $J_d$ and $K_d$ in Eqs.~\eqref{eq:BI} and~\eqref{eq:OBI} we use the symmetry that exchanges the subsystems $\BB$ and $\CC$.

\section{Optimal states for $J_2$ and $K_2$}
\label{app:optJ2}

Take inequalities $J_2$ and $K_2$ in Eqs.~\eqref{eq:BI} and \eqref{eq:OBI}.
We fix the local dimensions of Bob and Charlie to be two. Denote the Pauli matrices on one qubit by $X$, $Y$ and $Z$, and the Bell state by $\phi^+ = (\ket{00} + \ket{11})(\bra{00} + \bra{11})/2$.

When Alice has Hilbert space dimension two, 
the maximum of $J_2$ is $2 + \sqrt 2 \approx 3.4142$. This value can be attained with the state $\varrho = \phi^+_{\AA\BB} \otimes \ketbra{0}{0}_\CC$ and measurements
\begin{align}
    A_{1|1}^{\AA|\BB} & = (Z-X)/\sqrt 2,  & B_{1|1}^{\AA|\BB}  &= Z, \nn \\
    A_{1|2}^{\AA|\BB} & = (Z+X)/\sqrt 2, & B_{1|2}^{\AA|\BB} &= X, \nn \\
    A_{1|1}^{\AA|\CC} & = \mathds 1, & C_{1|1}^{\AA|\CC} &= \mathds 1, \nn \\
    A_{1|2}^{\AA|\CC} & = \mathds 1, & C_{1|2}^{\AA|\CC} &= \mathds 1\,,
\end{align}
corresponding to a maximal quantum value for a CHSH inequality between $\AA$ and $\BB$.
By reusing these measurements for Dave, a lower bound on the maximum for $K_2$ is $2 + \sqrt 2 + 2\approx 5.4142$, and can be attained with the same state and measurements, additionally setting
\begin{align}
A_{1|1}^{\AA|\DD} & =A_{1|1}^{\AA|\BB},
& D_{1|1}^{\AA|\DD}  =B_{1|1}^{\AA|\BB}, \nn \\
A_{1|2}^{\AA|\DD} & =A_{1|2}^{\AA|\BB},
& D_{1|2}^{\AA|\DD}  = B_{1|2}^{\AA|\BB}.
\end{align}

Consider now the case when Alice has a quantum system of dimension four.
Then the maximum $2+2 = 4$ of $J_2$ is attained by two copies of Bell states, 
$\varrho = \phi^+_{\AA\BB} \otimes \phi^+_{\AA\CC}$ and corresponding CHSH settings.
The maximum of $K_2$ is $2+2+2 = 6$, which can be achieved with two pairs of Bell states and CHSH settings, 
where party $\DD$ shares the measurement with either $\BB$ or $\CC$.

Alice having a quantum system of dimension three interpolates the previous two settings: The maximum of $J_2$ is $1.818(3) + 1.818(3) = 3.636(5)$.
The state and measurements attaining this value are found numerically with the see-saw algorithm described in Appendix~\ref{subsec:sewsaw}. 
The state is
\begin{align}
 -&0.53575\ket{000} 
  -0.36188\ket{100} 
  +0.04281\ket{101} \nn \\
 +&0.47786\ket{110} 
  -0.01869\ket{111} 
  -0.38836\ket{200} \nn \\
 +&0.04595\ket{201} 
  -0.44526\ket{210} 
  +0.01738\ket{211},
  \label{eq:seesaw-J}
\end{align}
and the measurements are for Alice
\begin{align}
A^{\AA|\BB}_{1|1} & = 
\begin{pmatrix}
\phantom{-}0.9987 & -0.0329 & -0.0153 \\
-0.0329 & \phantom{-}0.1785 & -0.3815 \\
-0.0153 & -0.3815 & \phantom{-}0.8228
\end{pmatrix}, \nn\\
A^{\AA|\BB}_{1|2} & = 
\begin{pmatrix}
\phantom{-}0.0013 & -0.0129 & -0.0340 \\
-0.0129 & \phantom{-}0.1266 & \phantom{-}0.3323 \\
-0.0340 & \phantom{-}0.3323 & \phantom{-}0.8721
\end{pmatrix}, \nn\allowdisplaybreaks\\
A^{\AA|\CC}_{1|1} & = 
\begin{pmatrix}
\phantom{-}0.0825 & -0.2013 & \phantom{-}0.1875 \\
-0.2013 & \phantom{-}0.9558 & \phantom{-}0.0411 \\
\phantom{-}0.1875 & \phantom{-}0.0411 & \phantom{-}0.9617
\end{pmatrix}, \nn\\
A^{\AA|\CC}_{1|2} & = 
\begin{pmatrix}
\phantom{-}0.2164 & \phantom{-}0.3013 & -0.2808 \\
\phantom{-}0.3013 & \phantom{-}0.8842 & \phantom{-}0.1079 \\
-0.2808 & \phantom{-}0.1079 & \phantom{-}0.8994
\end{pmatrix},
\label{eq:seesaw-J-meas-A}
\end{align}
and for Bob and Charlie, respectively
\begin{align}
B^{\AA|\BB}_{1|1} & = 
\begin{pmatrix}
\phantom{-}0.4999 & \phantom{-}0.5000 \\
\phantom{-}0.5000 & \phantom{-}0.5001
\end{pmatrix}, \nn\\
B^{\AA|\BB}_{1|2} & = 
\begin{pmatrix}
\phantom{-}1.0000 & \phantom{-}0.0001 \\
\phantom{-}0.0001 & \phantom{-}0.0000
\end{pmatrix}, \nn\\
C^{\AA|\CC}_{1|1} & = 
\begin{pmatrix}
\phantom{-}0.9862 & -0.1167 \\
-0.1167 & \phantom{-}0.0138
\end{pmatrix}, \nn\\
C^{\AA|\CC}_{1|2} & = 
\begin{pmatrix}
\phantom{-}0.3833 & -0.4862 \\
-0.4862 & \phantom{-}0.6167
\end{pmatrix}.
\label{eq:seesaw-J-meas-D}
\end{align}

When optimizing $K_2$, the seesaw outputs a lower bound on the maximum value of $1.683(7) + 1.912(9) + 1.912(9) = 5.509(6)$. This value is attained with the state
\begin{align}
    &0.38371\ket{011} 
  + 0.47320\ket{100} 
  - 0.44665\ket{101} \nn \\
 + &0.02641\ket{110} 
  - 0.01453\ket{111} 
  + 0.42595\ket{200} \nn \\
 + &0.49615\ket{201} 
  + 0.02392\ket{210} 
  + 0.01666\ket{211},
  \label{eq:seesaw-K}
\end{align}
and with measurements for Alice
\begin{align}
A^{\AA|\BB}_{1|1} & = 
\begin{pmatrix}
\phantom{-}0.9025 & -0.1982 & \phantom{-}0.2207 \\
-0.1982 & \phantom{-}0.0435 & -0.0485 \\
\phantom{-}0.2207 & -0.0485 & \phantom{-}0.0540
\end{pmatrix}, \nn\\
A^{\AA|\BB}_{1|2} & = 
\begin{pmatrix}
\phantom{-}0.8521 & \phantom{-}0.2378 & -0.2636 \\
\phantom{-}0.2378 & \phantom{-}0.0663 & -0.0736 \\
-0.2636 & \phantom{-}0.0736 & \phantom{-}0.0816
\end{pmatrix}, \nn\\
A^{\AA|\CC}_{1|1} & = 
\begin{pmatrix}
\phantom{-}0.0002 & -0.0123 & -0.0044 \\
-0.0123 & \phantom{-}0.8882 & \phantom{-}0.3149 \\
-0.0044 & \phantom{-}0.3149 & \phantom{-}0.1117
\end{pmatrix}, \allowdisplaybreaks\nn\\
A^{\AA|\CC}_{1|2} & = 
\begin{pmatrix}
\phantom{-}0.0002 & \phantom{-}0.0055 & \phantom{-}0.0116 \\
\phantom{-}0.0055 & \phantom{-}0.1857 & \phantom{-}0.3889 \\
\phantom{-}0.0116 & \phantom{-}0.3889 & \phantom{-}0.8141
\end{pmatrix}, \allowdisplaybreaks \nn\\
A^{\AA|\BB\CC}_{1|1} & = 
\begin{pmatrix}
\phantom{-}0.9994 & \phantom{-}0.0216 & -0.0116 \\
\phantom{-}0.0216 & \phantom{-}0.2250 & \phantom{-}0.4170 \\
-0.0116 & \phantom{-}0.4170 & \phantom{-}0.7756
\end{pmatrix}, \nn\\
A^{\AA|\BB\CC}_{1|2} & = 
\begin{pmatrix}
\phantom{-}0.0001 & \phantom{-}0.0095 & -0.0029 \\
-0.0095 & \phantom{-}0.9169 & \phantom{-}0.2758 \\
-0.0029 & \phantom{-}0.2758 & \phantom{-}0.0830
\end{pmatrix},
\label{eq:seesaw-K-meas-A}
\end{align}
for Bob and Charlie
\begin{align}
B^{\AA|\BB}_{1|1} & = 
\begin{pmatrix}
\phantom{-}0.0033 & -0.0569 \\
-0.0569 & \phantom{-}0.9967
\end{pmatrix}, \nn\\
B^{\AA|\BB}_{1|2} & = 
\begin{pmatrix}
\phantom{-}0.5609 & -0.4963 \\
-0.4963 & \phantom{-}0.4391
\end{pmatrix}, \nn\\
C^{\AA|\CC}_{1|1} & = 
\begin{pmatrix}
\phantom{-}1.0000 & \phantom{-}0.0000 \\
\phantom{-}0.0000 & \phantom{-}0.0000
\end{pmatrix}, \nn\\
C^{\AA|\CC}_{1|2} & = 
\begin{pmatrix}
\phantom{-}0.5000 & \phantom{-}0.5000 \\
\phantom{-}0.5000 & \phantom{-}0.5000
\end{pmatrix},
\end{align}
and the measurements for Dave
\begin{align}
D^{\AA|\BB\CC}_{1|1} & = 
\begin{pmatrix}
\phantom{-}0.9963 &         {-}0.0481 & \phantom{-}0.0244 &        {-}0.0016\\
        {-}0.0481 & \phantom{-}0.0036 &         {-}0.0028 &        {-}0.0269\\
\phantom{-}0.0244 &         {-}0.0028 & \phantom{-}0.5635 &        {-}0.0001\\
        {-}0.0016 &         {-}0.0269 &         {-}0.0001 & \phantom{-}0.5608
\end{pmatrix}, \nn\\
D^{\AA|\BB\CC}_{1|2} & = 
\begin{pmatrix}
\phantom{-}0.4520 &         {-}0.4964 &        {-}0.0063 &        {-}0.0227\\
        {-}0.4964 & \phantom{-}0.5472 &         {-}0.0279 & \phantom{-}0.0250\\
        {-}0.0063 &         {-}0.0279 & \phantom{-}0.5618 &        {-}0.0013\\
        {-}0.0227 & \phantom{-}0.0250 &         {-}0.0013 & \phantom{-}0.0011
\end{pmatrix}.
\label{eq:seesaw-K-meas-D}
\end{align}

\bigskip
\bibliography{obi}

\end{document}